\documentclass[11pt]{article}
\usepackage{moriond,epsfig}

\begin{document}
\vspace*{4cm}
\title{SPIN DEPENDENT STRUCTURE FUNCTIONS IN PHENIX}

\author{K. Oleg Eyser on behalf of the PHENIX collaboration}

\address{University of California, Riverside, CA 92521, USA}

\maketitle\abstracts{
  In the year 2005 run PHENIX has accumulated 3.8 pb$^{-1}$ integrated
  luminosity in longitudinally polarized $p+p$ collisions at
  $\sqrt{s}=200$~GeV. 
  Double helicity asymmetries of neutral pions lead to restrictions of
  the gluon polarization in the proton.
  Also, transverse single spin asymmetries of charged hadrons have been
  measured with comparable statistical accuracy of the 2002 run in
  considerably less time.
}

%===================================================================
%===================================================================
\section{Introduction}
The RHIC spin program at $\sqrt{s}=200$~GeV is largely devoted to the
measurement of the gluon polarization inside the proton.
In the past, we have learned from deep inelastic scattering (DIS) experiments
that the na\"ive picture of three quarks in the proton cannot explain the spin
structure properly.
In fact, the valence and sea quarks together carry less than 30\% of the
nucleon spin.\cite{a}
Other contributions have to arise from gluons and orbital angular momenta of
gluons and quarks.

While gluons only interact strongly, a proton collider like RHIC opens unique 
ways to determine their spin dependent structure functions.
Factorizaton allows us to separate the long range and short range parts in the
scattering process.
The short range, or hard scattering process, can be calculated in
next-to-leading order perturbative QCD (NLO pQCD) and carries spin
dependence in the partonic analyzing powers.
Long range interactions divide into initial and final state parts.
Fragmentation after the hard scattering process has been measured in DIS
experiments and can be parameterized in models.
The initial state part contains the nucleon structure functions and can be
deduced from experimental asymmetries by including the hard scattering
calculations and fragmentation functions.

Also, fragmentation functions and partonic analyzing powers point to promising
probes for the measurement of the gluon polarization.
Neutral pions at low and medium transverse momenta ($p_T$) are mainly produced
via gluon fusion and quark gluon scattering.
However well this might lead to the absolute value of the gluon polarization
$|\Delta g|$, gluon fusion inhibits a direct measurement of the sign of
$\Delta g$.
Direct photons, on the other hand, are dominated by quark gluon scattering and
do not contain further fragmentation.
This makes them an excellent, though extremely demanding, probe for 
$\Delta g$.

Transverse single spin asymmetries (SSA) are expected to vanish in
perturbative QCD calculations.
However, they have been observed in several experiments in the past 15
years.\cite{1,2}
%In order to explain this behaviour, another chiral odd function must be
%introduced.
Possible mechanisms have been proposed in order to describe this behaviour.
These include spin dependent fragmentation functions (Collins
function),\cite{3} asymmetric intrinsic transverse momentum
distributions (Sivers effect),\cite{4} and quark gluon field interference in
higher twist calculations.\cite{5}
The strength of these effects is still largely unknown, and their
measurement leads to a deeper understanding of the transverse quark
distribution $\delta q$ and, hence, the relativistic nature of the nucleon.

%===================================================================
%===================================================================
\section{Measurement}
PHENIX is a multi purpose detector with dedicated systems for heavy ion and
spin physics programs.\cite{b}
The detector setup includes the central arm at mid-rapidity ($|\eta| < 0.35$),
two muon arms at large rapidities ($1.2 < |\eta| < 2.4$), and additional global
detectors.

The total azimuthal coverage of the central arm is $\Delta\varphi=\pi$ divided
into two symmetric parts.
An axial magnetic field bends charged particles traversing the central region
before their tracks are detected in the drift chamber (DC).
A ring imaging Cherenkov (RICH) system allows the identification of electrons
and other charged particles.
Additional layers of pad chambers (PC) can be used for track matching after
the DC and the RICH.
Outermost are the lead scintillator (PbSc) and lead glass (PbGl)
electro-magnetic calorimeters which complement each other in terms of timing
and energy resolution.
Several layers of absorbers, muon tracking stations, and muon identification
stations comprise the muon arms.

The global detectors are paired sets of beam-beam counters (BBC) and
zero-degree calorimeters (ZDC).
The BBC is used for the minimum bias trigger.
Also, this scaler is used for other triggered data sets in combination with
the ZDC for the determination of relative luminosity $R$ of bunches with
different polarization directions in the accelerator.
The uncertainty in the determination of relative luminosities in 2005 was
$\Delta R < 10 ^{-4}$.
The ZDC is crucial for the local polarimetry in which a very forward neutron
asymmetry is used to confirm the alignment of the polarization vector in the
vertex region.\cite{c}
This is especially important during longitudinal polarization measurements
when the polarization is rotated in the PHENIX interaction region.
The polarization vector has been found to be 98\% or more longitudinally
aligned with only small residual transverse components.

In 2005 RHIC has provided about three months of $p+p$ collisions at
center-of-mass energies $\sqrt{s}=200$~GeV leading to an integrated luminosity
of 3.8~pb$^{-1}$ with longitudinal polarization.
Additionally, 0.15~pb$^{-1}$ with transverse polarization were collected.
The average polarization in both beams was P = 47\% with a 20\% relative error
due to the not finally calibrated Carbon polarimeters.

%===================================================================
%===================================================================
\section{Neutral Pion Double Helicity Asymmetries}
Neutral pions in PHENIX are detected in the central arms with the
electro-magnetic calorimeters with a high-$p_T$ photon trigger.
Particle identification uses a minimal photon energy and shower profile in the
calorimeters, a charge veto, and an energy asymmetry in the combined photon
pairs.
The $\pi^0$ cross section in 2005 was found to be in good agreement with
NLO pQCD calculations and previous results from 2003\cite{c} and 2004.

Double helicity asymmetries are calculated from:
\begin{equation}
  A_{LL}=\frac{\sigma_{++}-\sigma_{+-}}{\sigma_{++}+\sigma_{+-}},
\end{equation}
where $\sigma$ is the cross section with respect to parallel (++) and/or
antiparallel (+-) helicity states of the colliding protons.
The measured quantities are yields $N$ which are obtained from the cross
sections by integrating over detector acceptances and luminosities $L$.
In the ratio:
\begin{equation}
  A_{LL} = \frac{1}{P_B\cdot P_Y}\cdot\frac{N_{++}-R\cdot N_{+-}}{N_{++}+R\cdot N_{+-}},
\end{equation}
everything but the relative luminosity $R=L_{++}/L_{+-}$ cancels.
Also, we have to take the polarizations $P_B$ and $P_Y$ of both beams into account.

%===================================================================
\begin{figure}
  \begin{center}
    \psfig{figure=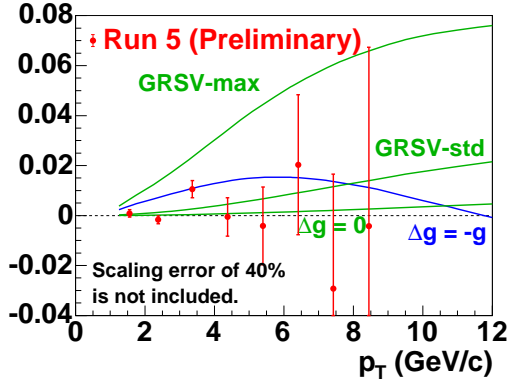,width=0.45\textwidth}
  \end{center}
  \caption{\label{fig:ALL}
    Double helicity asymmetry $A_{LL}$ of neutral pions at mid-rapidity.
    The data are compared to four different assumptions of the gluon
    polarization (GRSV-max $\Delta g=g$, GRSV standard, $\Delta g = 0$, and
    $\Delta g = -g$).%\cite{d}
  }
\end{figure}
%===================================================================

Figure \ref{fig:ALL} shows the PHENIX results of double helicity asymmetries for
neutral pions from 2005.
An additional scaling error of 40\% arises from twice the uncertainty of the
beam polarization.
The data is compared to calculations for different assumptions of the gluon
polarization.\cite{d}
The data disagree with the maximal gluon polarization scenarios 
($\Delta g = g$ and $\Delta g = -g$), but are still consistent with GRSV
standard and $\Delta g=0$.

%===================================================================
%===================================================================
\section{Single Spin Asymmetries}
Single spin asymmetries are obtained in double polarized collisions in RHIC by
averaging or summing over different polarization states in one beam.
This can be done in a single accelerator fill, which consists of several
bunches of both polarization directions.
The residual polarization after averaging is estimated from relative
luminosities and is small compared to the original polarization in separate
bunches.

Charged particles are tracked in the DC.
In combination with the central magnetic field their scattering angles and
momenta can be determined.
The track quality is defined by certain cuts including signals in several
layers of the detector.
Major background sources for fake high-$p_T$ tracks are conversion
electrons and decay particles with short lifetimes.
Those are removed by Cherenkov information and track matching in the outer
detector layers (PC).

SSA show up as an azimuthal cosine modulation of yields,
leading to a specific left-right signature.
This can be understood by the projection of the polarization vector $\vec{P}$
(as prepared in the accelerator frame) onto the direction perpendicular to the
scattering frame.
Respective asymmetries therefore are dependent on the position of the
detector, which is included in calculable acceptance terms $c_{left}$ and
$c_{right}$ for detectors on the left and right sides.
In order to remove differences in efficiencies and relative luminosities from
the asymmetries, the ratio:
\begin{equation}
\frac{N^\uparrow_{left}  \cdot N^\downarrow_{right}}
     {N^\downarrow_{left}\cdot N^\uparrow_{right}}
     =
     \frac{(1+A_N\cdot c_{left})\cdot(1+A_N\cdot c_{right})}
	  {(1-A_N\cdot c_{left})\cdot(1-A_N\cdot c_{right})}
\end{equation}
combines yields $N$ of different polarization directions ($\uparrow$
polarization up, $\downarrow$ polarization down) and detector hemispheres
($left$ and $right$ with respect to the impinging beam).
This ratio can be transformed into an extended square root formula for the
analyzing power $A_N$.

%===================================================================
\begin{figure}
  \begin{center}
    \psfig{figure=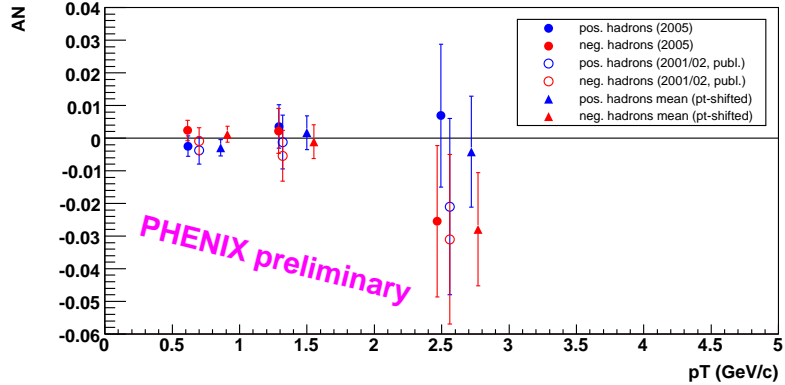,width=0.7\linewidth}
  \end{center}
  \caption{\label{fig:AN}
    Single spin asymmetries for charged hadrons at mid-rapidity.
    Data from 2005 are compared to previously published data from 2002.%\cite{6}
    Additional scaling errors of 20\% (2005 data) and 30\% (2002 data) have to
    be added.
  }
\end{figure}
%===================================================================

Figure \ref{fig:AN} shows the results from 2005 for inclusive charged hadron $A_N$
and compares them to the published results of 2002.\cite{6} with roughly the
same statistical accuracy.
They agree well with each other, no evidence for a finite value of $A_N$
can be found at $p_T < 5$~GeV/c.

%===================================================================
%===================================================================
\section{Summary and Outlook}
The neutral pion cross section shows that factorization can be utilized at
RHIC energies and that the hard scattering process is well understood in NLO
pQCD.
Also, they have pointed to a small gluon polarization at small $x$.
No finite SSA have been found in PHENIX at mid-rapidity.

Direct photons will lead not only to the absolute value of $|\Delta g|$ but
also to its sign.
The separation of direct photons from background is demanding and the yields
are small.
Asymmetries have so far not been obtained.
Other interesting probes include $J/\Psi$ to test gluon fusion dominance in
heavy flavor, the spin transfer of weak decaying, and therefore self-analyzing,
$\Lambda$ baryons, and helicity correlated differences in jet intrinsic
$k_T$.

%===================================================================
%===================================================================

\end{document}